\documentclass[%
 reprint,
 amsmath,amssymb,
 aps
,
floatfix,
]{revtex4-2}

\usepackage{graphicx}

\usepackage{amsmath}
\usepackage{dcolumn}

\usepackage{bm}
\usepackage{svg}
\usepackage{hyperref}
\usepackage{xcolor}
\usepackage[normalem]{ulem}
\begin{document}


\title{Electronic Structure and Vibrational Stability of Copper-substituted Lead Apatite (LK-99)}

\author{J. Cabezas-Escares}
\author{Nicolás F. Barrera}%

\affiliation{%
Departamento de F\'isica, Facultad de Ciencias, Universidad de Chile, Santiago, Chile
}%
\affiliation{
Center for the Development of Nanoscience and Nanotechnology (CEDENNA), Santiago, Chile
}%

\author{Robert H. Lavroff}
\affiliation{
Department of Chemistry and Biochemistry, University of California Los Angeles, Los Angeles, California, USA
}%
\author{Anastassia N. Alexandrova}
\affiliation{
Department of Chemistry and Biochemistry, University of California Los Angeles, Los Angeles, California, USA; Department of Materials Science and Engineering, University of California Los Angeles, Los Angeles, California, USA; and California Nanoscience Institute (CNSI), Los Angeles, California, USA
}%

\author{C. Cardenas}
\author{F. Munoz}
\email{fvmunoz@gmail.com}
\affiliation{
Departamento de F\'isica, Facultad de Ciencias, Universidad de Chile, Santiago, Chile
}%
\affiliation{
Center for the Development of Nanoscience and Nanotechnology (CEDENNA), Santiago, Chile
}%

\date{\today}

\begin{abstract}
Two recent preprints in the physics archive (arXiv) have called attention as they claim experimental evidence that a Cu-substituted apatite material (dubbed LK-99) exhibits superconductivity at room temperature and pressure. If this proves to be true, LK-99 will be a ``holy grail" of superconductors. In this work, we used Density Functional Theory (DFT+U) calculations to elucidate some key features of the electronic structure of LK-99. We find two different phases of this material: \textit{(i)} a hexagonal lattice featuring metallic half-filled and spin-split bands, a nesting of the Fermi surface, a remarkably large electron-phonon coupling, but this lattice is vibrationally unstable. \textit{(ii)} a triclinic lattice, with the Cu and surrounding O distorted. This lattice is vibrationally stable and its bands correspond to an insulator. In a crystal, the Cu atoms should oscillate between equivalent triclinic positions, with an average close to the hexagonal positions. We discuss the electronic structure expected from these fluctuations and if it is compatible with superconductivity. 

\end{abstract}

\maketitle


\section{\label{sec:level1}Introduction}

Since the discovery of superconductivity in 1911 by H. Kamerlingh Onnes \cite{onnes1911superconductivity,onnes1911discovery}, the phenomenon has fascinated scientists. Many efforts have been made to find materials capable of this property under conditions of temperature and pressure that allow practical applications. This search has resulted in the discovery of materials such as the conventional, or BCS, superconductor $\mathrm{MgB}_{2}$ with $T_{c}=39$~K \cite{nagamatsu2001superconductivity} or unconventional (beyond BCS)  superconductor Y-Ba-Cu-O \cite{wu1987superconductivity} ($T_{c}=93\,K$). Nevertheless, a room-temperature and pressure superconductor remains elusive.

The material dubbed LK-99, an apatite-like crystal with the approximate formula unit CuPb$_9$(PO$_4$)$_6$O, has been in the spotlight of the condensed matter physics community, as two preprints articles claim it has a superconducting critical temperature over 400 K at atmospheric pressure \cite{lee2023,lee2023b}. If these claims prove to be true, this discovery could be one of the major advancements in the field of superconductivity.\\

{ Most of the recent literature on this material points against it as a superconductor, with several apparently successful replication attempts, capable of reproducing results such as X-ray patterns, but without exhibiting superconductivity\cite{Kumar2023Absence,zhang2024ferromagnetism,wang2023ferromagnetic,singh2024electromagnetic}. It now seems generally accepted that the abrupt drop in electrical resistance below $T=104^\circ$C is originating from CuS$_2$ impurities\cite{Jain2023,JAIN20234118,Liu2023phases,ZHU20234401,Habamahoro_2024,Lei2024char,CHO202422}. Interestingly, one study found that CuS$_2$ impurities cannot reproduce the colossal magnetoresistance at room temperature found in LK-99\cite{Chen2024}, so this material could be promising as a magnetic sensor.
An experimental study describes a strange metal phase in this system, without discarding the observation of the Meissner effect\cite{wang2024observation}. Another study suggests the possibility of interfacial superconductivity at grain boundaries\cite{abramian2023some}. A theoretical paper supports this last idea, with a model of excitonic superconductivity\cite{CAO2023107293}. Recent theoretical studies have focused on understanding the defects contained in this material or its parent compounds\cite{Toriyama2024,tao2023cu}, and the structural stability of the crystal structure associated with LK-99\cite{Shen2024,Kim_2024,Jun2023}. Even one theoretical study has shown that this material could be one of the few that hosts Weyl fermions while breaking the time-reversal symmetry\cite{Brass2014}.  

}

In this article, we explore the possibility of LK-99 being such a superconducting material.

We will begin by discussing, in Sec.~\ref{sec:lead} the main properties and crystal structure of lead apatite. Then, in Sec.~\ref{sec:lk99} we will elaborate on the possible crystal structure and magnetic order of the so-called LK-99 system, as well as its electronic structure. This is the structure we have seen in other theoretical studies \cite{griffin2023origin,kurleto2023pbapatite,LAI2024,Si2023,si2023pb10xcuxpo46o,si2023pb10xcuxpo46o, sun2023metallization, jiang2023pb9cupo46oh2, korotin2023electronic, bai2023ferromagnetic, hao2023firstprinciples, yue2023correlated, tao2023cu, witt2023superconductivity}, and we will show it is vibrationally unstable. In Sec.~\ref{sec:triclinic} we will introduce a related crystal structure without imaginary phonon frequencies, but it is a band insulator. We will discuss our findings in Sec.~\ref{sec:discussion}. Finally, we will close this contribution with our conclusions, in Sec.~\ref{sec:con}

\section{\label{sec:lead}Lead-apatite}

\begin{figure}[h]
    \centering
    \includegraphics[width=\columnwidth]{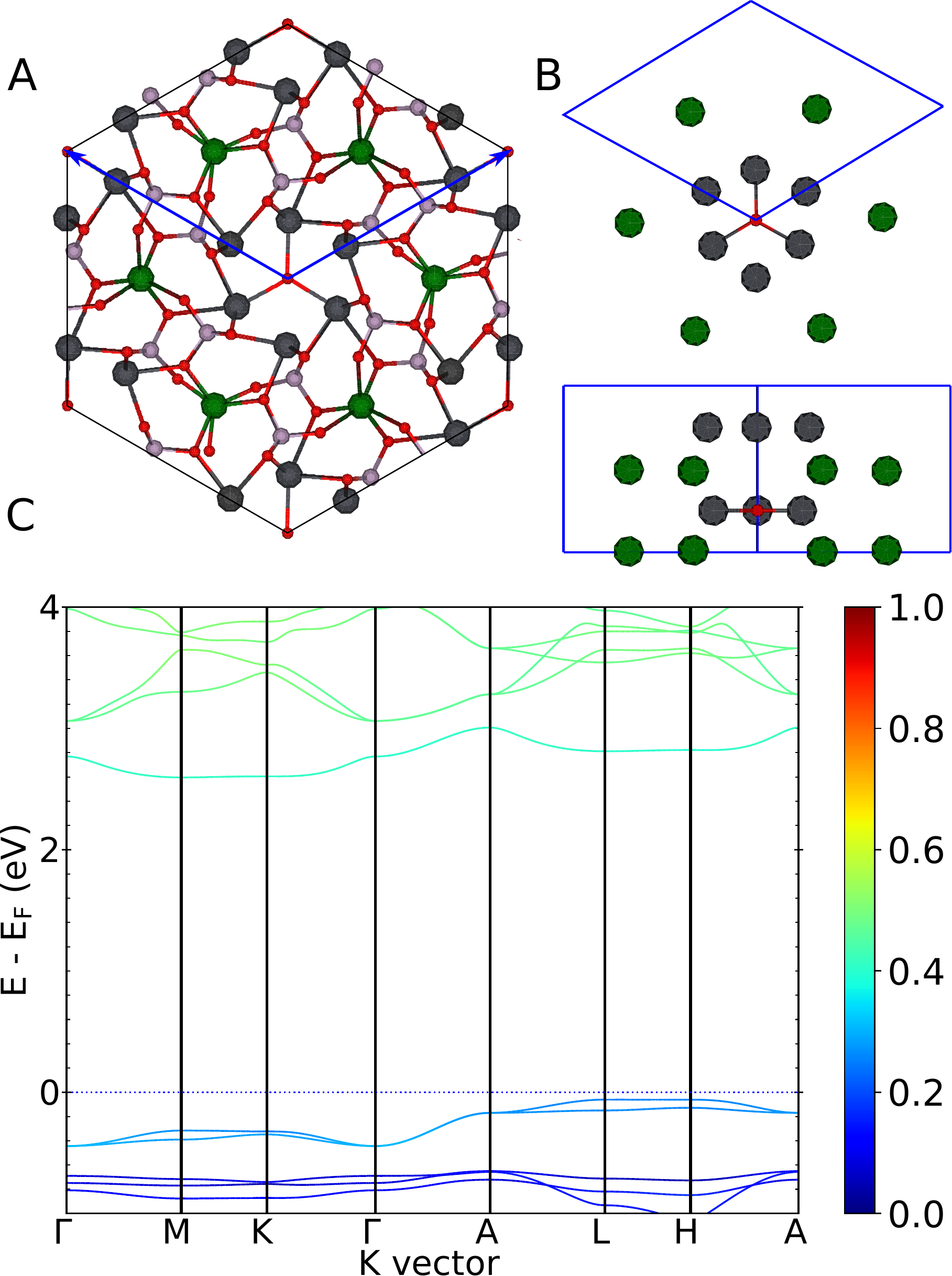}
    \caption{(a) In-plane view of the crystal structure of lead-apatite, Pb$_{10}$(PO$_4$)$_6$O. Nonequivalent Pb atoms are colored gray and dark green. O is red and P is pink. The lattice vectors are blue arrows. (b) Top and side view of the two hexagonal-like patterns formed by Pb atoms. The inner hexagonal pattern has three Pb atoms in different layers; these layers are not equivalent, since one of them has an O at its center. (c) Band structure of lead apatite. The color intensity reflects the projection of the wave functions into Pb atoms (\textit{i.e.} within its Wigner radius).}
    \label{fig:Pb-geo}
\end{figure}

Lead apatite materials have a hexagonal lattice with space group P6$_3$/m and formula unit Pb$_{10}$(PO$_4$)$_6$X$_2$, with X=Cl, OH, F, Br \cite{bhatnagar1968}. However, the apatite-like LK-99 phase has a slightly different composition, Pb$_{10-x}$Cu$_x$(PO$_4$)$_6$O. There are reports of a related (Ca-based) oxyapatite crystal with the desired composition \cite{alberius1999}. Its crystal structure is very close to other apatite materials. It has a hexagonal lattice with space group number \#174 and point group P$\bar{6}$. We used this geometry as our basis to computationally characterize Pb apatite, Pb$_{10}$(PO$_4$)$_6$O. We obtain lattice parameters $a=10.00$ \AA, $c=7.44$ \AA, {slightly larger than experimental reports} of similar systems \cite{hopwood2016,KrivovichevBurns2003}. {Theoretical works with similar methods show a close\cite{LAI2024}, slightly smaller \cite{Jun2023}, or even larger cells\cite{YANG2023}. However, after including Cu atoms, the differences in lattice parameters among theoretical studies are minimal.}  The geometry used is shown in Fig.~\ref{fig:Pb-geo}a, with Pb atoms forming hexagonal patterns, as shown in Fig~\ref{fig:Pb-geo}b. This  Pb apatite is predicted to be an insulator, as can be seen from its band structure diagram (see Fig.~\ref{fig:Pb-geo}c). A more sophisticated hybrid XC-functional, HSE06, gives a band gap of $\sim 3.8$ eV, see Section~\ref{sec:methods}. The conventional lead hydroxyapatite, Pb$_{10}$(PO$_4$)$_6$(OH)$_2$, has a similar band structure, but with a larger band gap.

\section{\label{sec:lk99}LK-99: Hexagonal lattice}

\subsection{Crystal Structure}
\begin{figure}[ht]
    \centering
    \includegraphics[width=0.7\columnwidth]{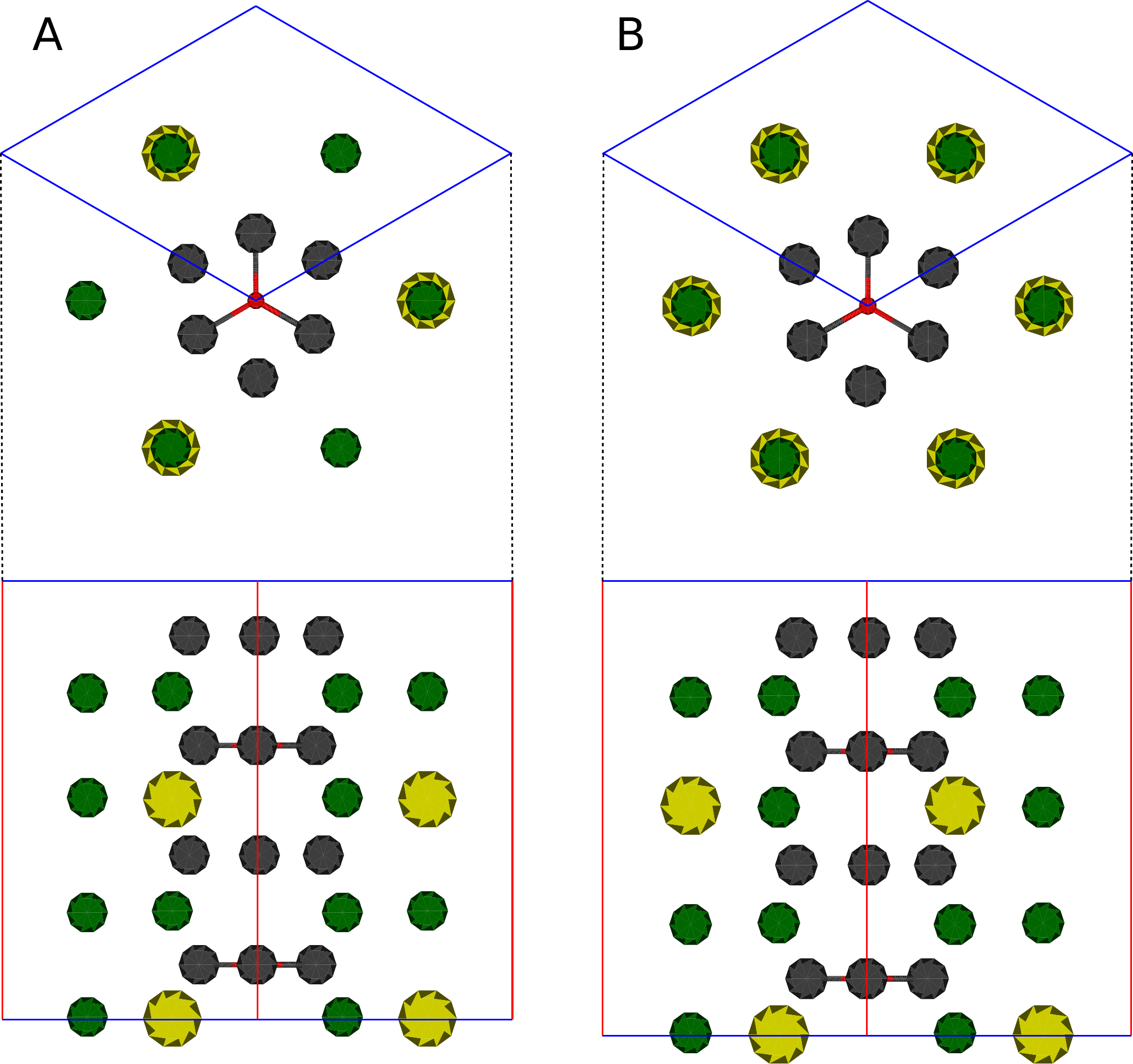}
    \caption{Possible LK-99 atomic structures. The unit cell of lead apatite was duplicated along the $c$-axis. To keep the figure as simple as possible P and most O atoms are omitted. The Cu atoms are yellow and large. Panels (a), (b) show two possible stacking sequences, A and B with Cu atoms forming a triangular or hexagonal sub-lattice.}
    \label{fig:cu1-geo}
\end{figure}

According to Lee \textit{et al.} \cite{lee2023,lee2023b}, the LK-99 phase has a unit formula Pb$_{10-x}$Cu$_x$(PO$_4$)$_6$O, with $0.9<x<1.1$, with Cu atoms replacing a specific Pb sublattice, the green atoms in Fig.~\ref{fig:Pb-geo}a-b. For simplicity, we set $x=1$, \textit{i.e.} a single Cu atom per unit cell. This substitution  Pb $\to$ Cu implies an odd number of electrons per unit cell, suggesting a metal with  a spin-split ground state, or a doubling of the unit cell. Fig.~\ref{fig:cu1-geo} shows two possible arrangements when doubling the unit cell along the $c$-axis, denoted stacking A and B. In stacking A (B) the Cu atoms form a triangular (hexagonal) sublattice. The space group of stacking A (B) is \#174 (\#143)

Overall six possible arrangements are considered, stacking sequences A and B, each with a ferromagnetic (FM), antiferromagnetic (AFM), or non-magnetic (NM) ground state. { In the AFM stacking, the in-plane spins are parallel with an antiparallel out-of-plane order} The relative energies, without including any Hubbard-like $U$ electron-electron repulsion correction, are given in Table~\ref{tab:energies}. It is evident that a magnetic order is preferred over NM. Also, the stacking A (triangular Cu sublattice) is slightly more stable than the stacking B. The energies of all different magnetic orders being almost the same is a first hint that the Cu atoms do not form extended states. A caveat: the designation as FM and AFM is questionable due to the absence of an effective exchange interaction among Cu atoms preventing a long-range magnetic order (\textit{i.e.} in a Heisenberg model, the exchange parameters should be practically zero){, however, is relevant to include spin-polarization to correctly describe the system}.

\begin{table}[ht]
    \centering
    \caption{Relative energies of different stacking and magnetic orders of Pb$_{9}$Cu$_x$(PO$_4$)$_6$O. The minimum energy configuration is taken as 0. These are the only results with the Hubbard-like term $U=0$ in the article. The units are per formula unit (f.u.), eV/f.u. }
    \label{tab:energies}

    \begin{tabular}{ccccc}
        \hline\hline
        Stacking/spin && A          && B \\\hline\hline
        NM            && 0.13 && 0.17\\
        FM            && 0.00 && 0.02\\
        AFM           && 0.00 && 0.04\\\hline\hline
    \end{tabular}
\end{table}

We tested the effect of a Hubbard-like term $U=2.0$ eV on the relative energy of the stacking A for the NM and FM orders, and the results are similar to those of Table~\ref{tab:energies}, with the NM being 0.18 eV/f.u. higher in energy than the FM order. Such larger stability of a spin-split state is expected from adding electron-electron interaction.


\subsection{Electronic properties}

\begin{figure}[ht]
    \centering
    \includegraphics[width=\columnwidth]{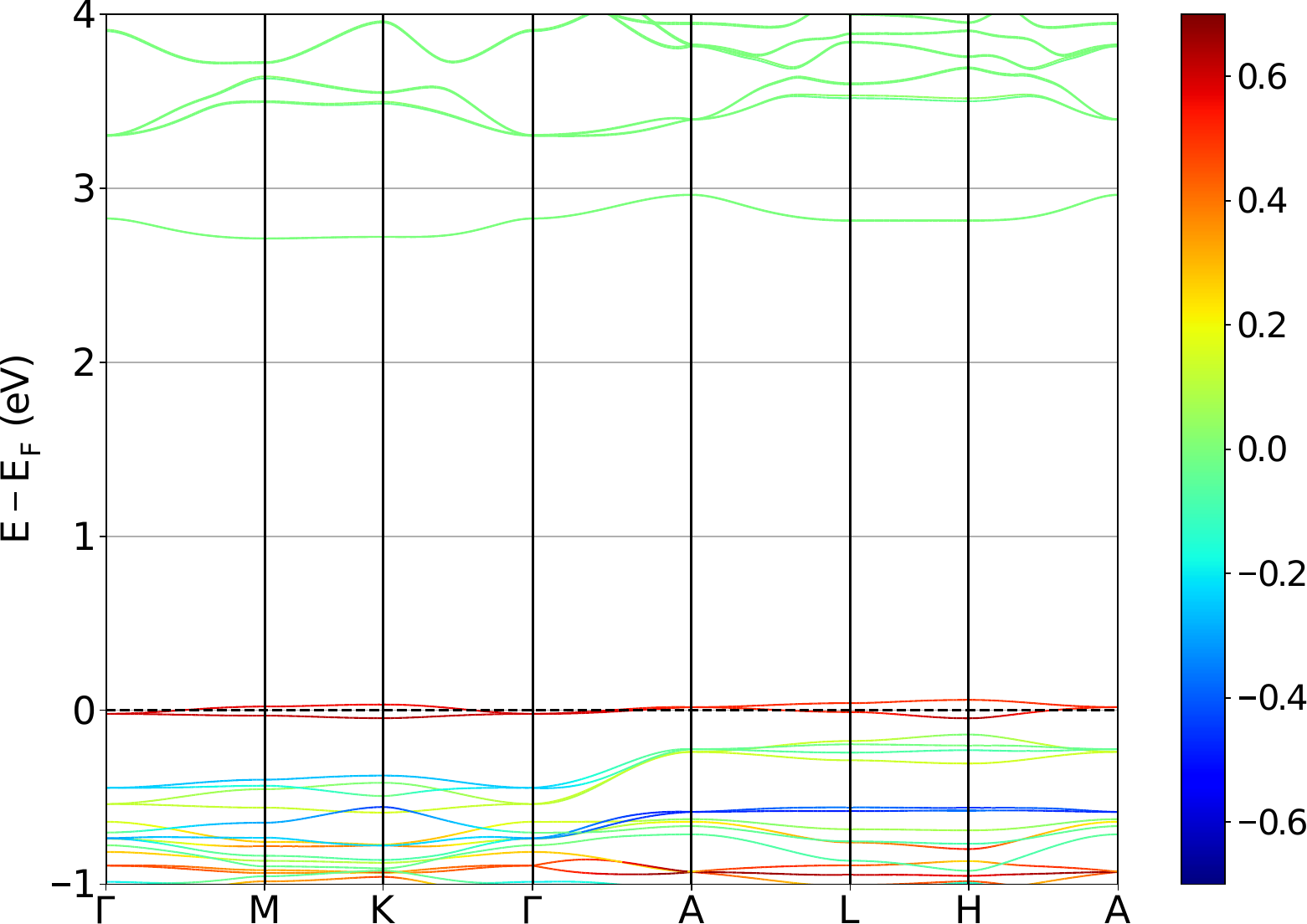}
    \caption{Band structure of Pb$_{9}$Cu$_1$(PO$_4$)$_6$O with stacking A and a FM ground state. The color scale represents the projection of wave functions onto the Cu atom, where a positive (negative) value denotes the spin value.}
    \label{fig:bands-Cu-FM}
\end{figure}

\begin{figure}
    \centering
    \includegraphics[width=\columnwidth]{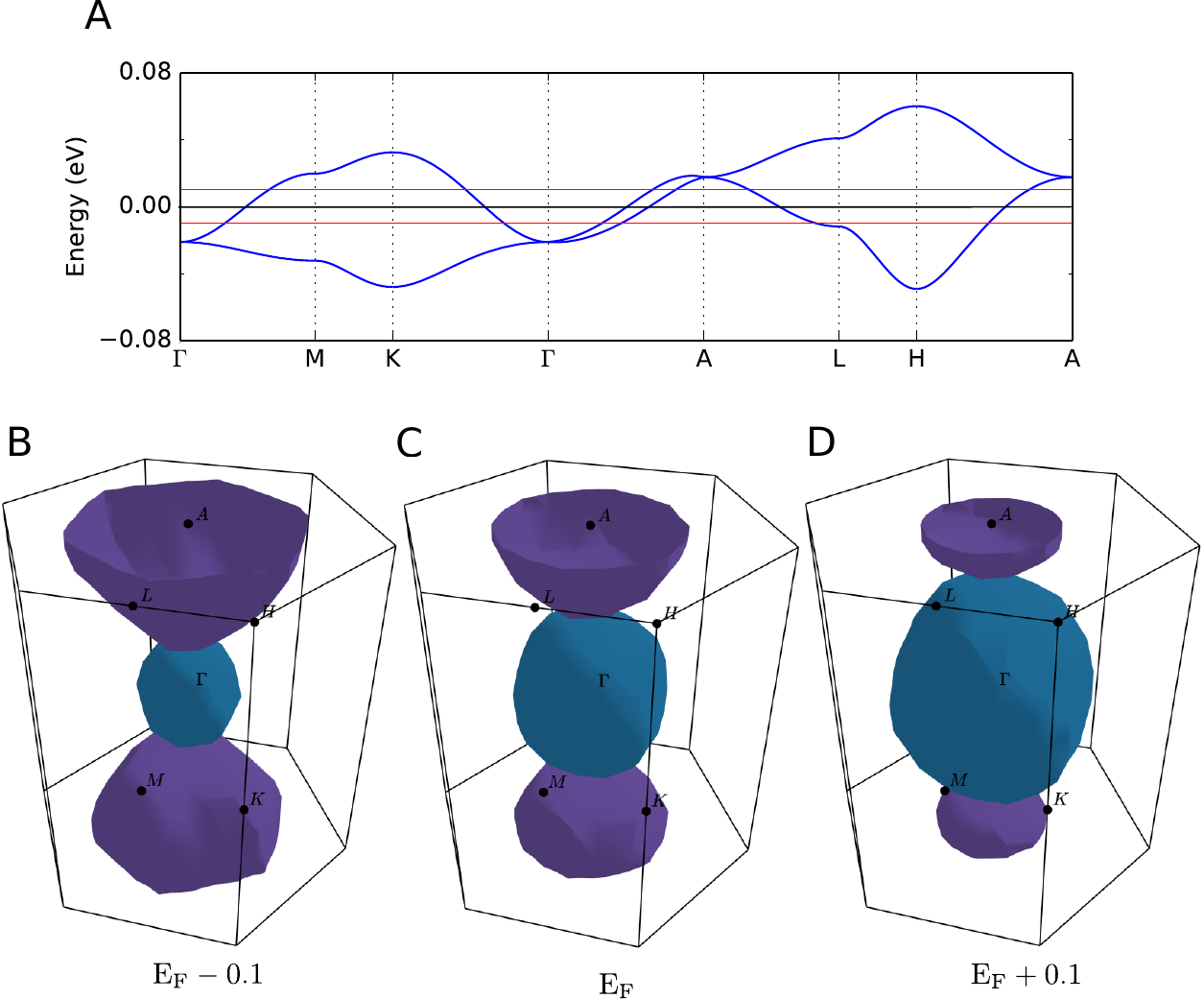}
    \caption{(a) Zoom-in to the bands at the Fermi level. (b-d) Isosurfaces of the bands at $E_F-0.01$ eV ($E_F$ is the Fermi energy), $E_F$, and  $E_F+0.01$. Different bands have different colors, particularly the two spherical-like surfaces, centered and $\Gamma$ and A, are { apparently nested}.}
    \label{fig:fermisurface}
\end{figure}

Provided the larger stability of the stacking A (triangular Cu sublattice, see Fig.~\ref{fig:cu1-geo}), we calculated its band structure. In the NM and FM cases, the unit cell of Pb$_{9}$Cu$_1$(PO$_4$)$_6$O suffices to describe the system (\textit{i.e.} no supercell is needed). Fig.~\ref{fig:bands-Cu-FM} shows the band structure of the FM order. Two Cu d-orbitals form almost flat bands at the Fermi level: their width is $\approx 0.1$ eV, see Fig.~\ref{fig:fermisurface}a. These bands are half-filled, and the system is metallic. These bands are very similar to other reports, with a $U=4-7$ eV, with and without spin-orbit interaction \cite{griffin2023origin,kurleto2023pbapatite,LAI2024,Si2023,si2023pb10xcuxpo46o,jiang2023pb9cupo46oh2,bai2023ferromagnetic, hao2023firstprinciples, Si2023}{, see Appendix~\ref{sec:varying-u} for the effect of changing the value of $U$}. 

The Fermi surface and other isovalues of the bands are shown in Fig.~\ref{fig:fermisurface}b-d. Here a nesting of the Fermi surface seems likely: two bands have an almost spherical shape, but one is centered at $\Gamma$ and the other at A. Fermi nesting is often associated with a charge density wave \cite{Dugdale_2016} as suggested by the original preprint attempting to explain the superconductivity of LK-99 \cite{lee2023b}. However, the link between nesting and a charge density wave is the Peierls distortion or a related mechanism \cite{nesting}. This mechanism can induce some other effects, including an actual change in geometry, such as dimerization. In fact, there are theoretical arguments to expect a reduction in the symmetry of the system \cite{hlinka2023possible}. {A deeper discussion, including the possible nesting vector, is provided in Appendix \ref{sec:nesting}.}

\begin{figure}[h]
    \centering
    \includegraphics[width=0.8\columnwidth]{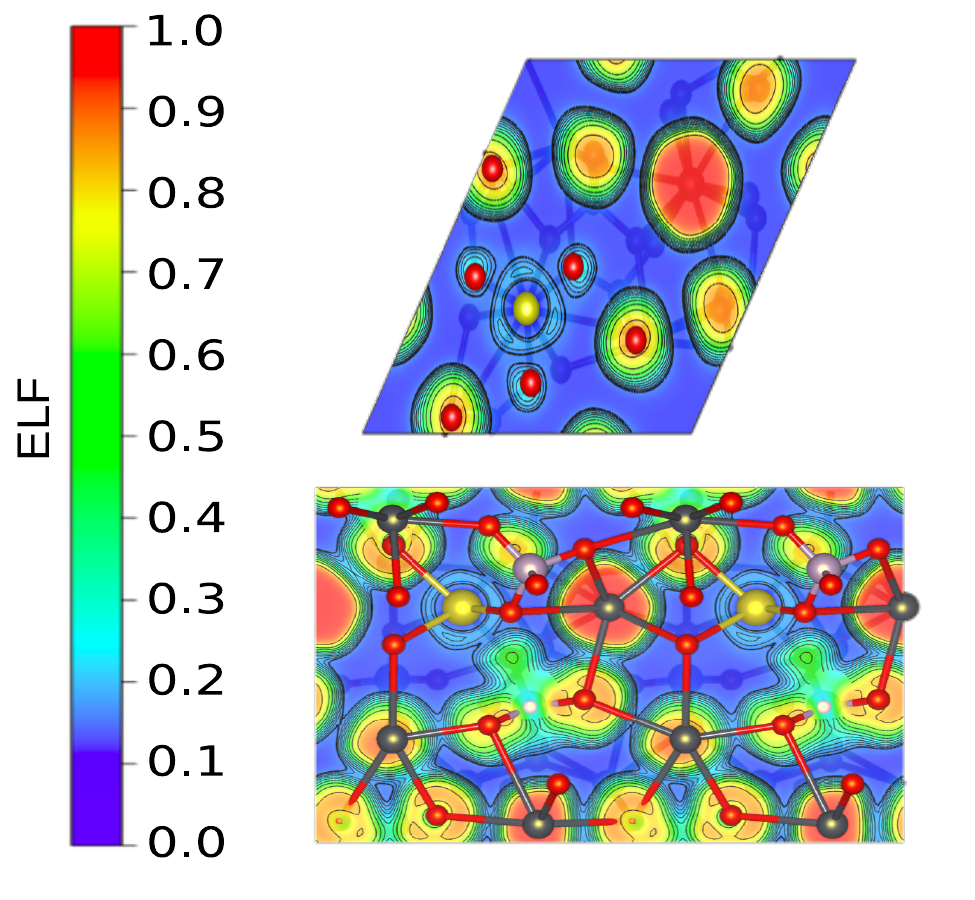}
    \caption{Contour plots of the ELF (dimensionless) on planes that cross the Cu atoms and are parallel to (1,0,0) (bottom)  and (0,0,1) (top). The coloring of the atoms follows Figs.~\ref{fig:Pb-geo}, \ref{fig:cu1-geo}, \textit{i.e.} the Cu atom is yellow.}
    \label{fig:elf2}
\end{figure}

The electron localization function (ELF) is large in those regions of space where it is likely to find electron pairs of opposite spin \cite{becke1990}. Hence, ELF is large in regions associated with covalent chemical bonds, lone pairs of electrons, and the inner shells of atoms. Contour plots of ELF along planes containing the Cu atoms, see Fig.~\ref{fig:elf2}, suggest that Cu is not strongly covalently bonded to its neighbor oxygen atoms.  {Indeed, the shape of the ELF around Cu resembles the one of atoms confined in wells \cite{novoa2019,robles2021}. This observation is consistent with the small hybridization d and s orbitals of Cu with the p orbitals of the oxygen, as shown by the angular momentum decomposition of the density of states of the Cu atom (see Fig. \ref{fig:DPOS_Cu_LK99_sum} in Appendix~\ref{sec:PDOS}).  A better understanding of the bonding of Cu may benefit from including static correlation in the bonding analysis, such as in the hybridization function \cite{Gull2011,Bahlke2021}, but this is beyond the scope of this study.}

Interestingly, Belli \textit{et al.} \cite{errea2021} found an empirical positive correlation between the critical temperature of high-pressure-hydrogen-based superconductor and what they called the networking value, $\phi$, of the ELF. $\phi$ is  \textit{``the highest value of the ELF that creates an isosurface spanning through the whole crystal in all three Cartesian directions"}. If these criteria would apply to LK-99, its critical temperature would be less than 50 K (see Fig. 4 in  \cite{errea2021}). It is worth noting that the quality of the ELF is only as good as its underlying DFT calculations, so its interpretation should be taken with a grain of salt.


\subsection{\label{sec:epc}Phonons and electron-phonon coupling}

\begin{figure}
    \centering
    \includegraphics[width=\columnwidth]{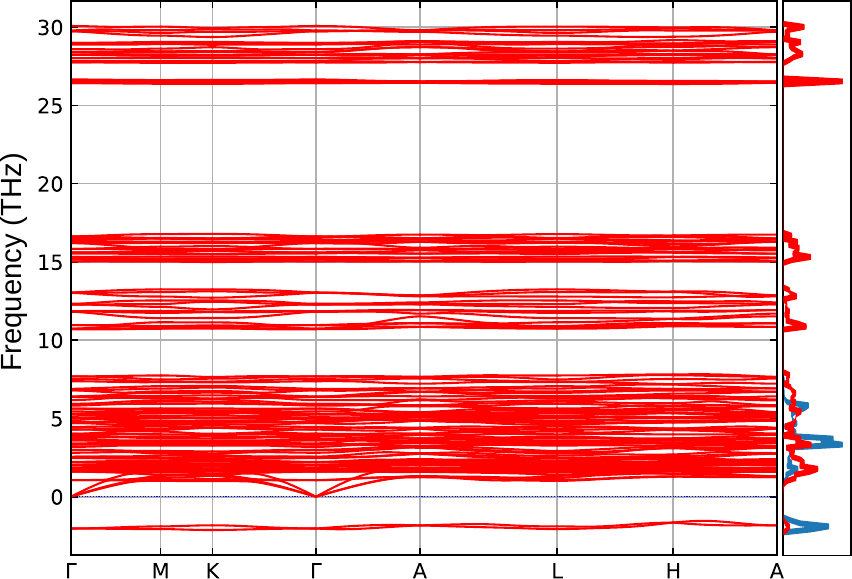}
    \caption{Phonon bandstructure of the arrangement A (triangular Cu sublattice, space group \#174). {The left panel is the total (red) and partial (Cu projection, gray) phonon density of states.}}
    \label{fig:phonon-hexagonal}
\end{figure}

{At the harmonic level,} the phonon band structure of arrangement A (see  Fig.~\ref{fig:phonon-hexagonal}) has well-defined groups of bands, including two completely imaginary branches. These branches indicate that the geometry with group \#174 is dynamically unstable, { at least under the conditions of our calculations}. In the next section, Sec.~\ref{sec:triclinic}, we will explore a related but vibrationally stable structure. { Anharmonic effects were studied elsewhere \cite{Kim_2024}, and we will discuss them in Sec.~\ref{sec:what-crystal}}

Other features of the phonons include \textit{(i)} the absence of dispersive phonons, indicating that the system is constructed from molecular-like blocks, and \textit{(ii)} the displacement of Cu atoms is confined to low-energy phonons (up to 30 meV), falling within the energy width of the Cu bands at the Fermi level. More energetic modes involve the rearrangement of O atoms surrounding the Cu.

Ignoring the imaginary frequencies, it could be instructive to have { a crude idea about } electron-phonon coupling (EPC) of the system. The degenerate flat bands at the Fermi level imply a potentially large EPC. Its evaluation with DFT is highly expensive, especially for such a large unit cell, and EPC within GGA$+U$ is not included in the DFT codes known to us. Nevertheless, we have adopted a much simpler approach to obtain insights on how large the EPC could be.

According to the isotropic Migdal-Eliashberg theory \cite{RevModPhys.89.015003}, the overall EPC is a weighted integral of the $\alpha^2F(\omega)$ function, which adds all the electron-phonon matrix elements ($|g_{m n \nu}(\mathbf{k},\mathbf{q})|^2$)  compatible with momentum and energy conservation at the Fermi level. Explicitly, 
\begin{equation}
    g_{mn\nu}(\mathbf{k},\mathbf{q}) = \langle u_{m,\mathbf{k+q}}|\Delta_{\mathbf{q\nu}}v^{KS}|u_{n,\mathbf{k}}\rangle,\label{eq:gnm}
\end{equation}
where $\mathbf{k}, \mathbf{q}$ refers to electron and phonon momentum, respectively. $m,n$ are electronic band indexes and $\nu$ is a phonon index. $u_{n\mathbf{k}}(r)$ is the lattice-periodic part of the wavefunction and $v^{KS}$ is the  Fourier transformed Kohn-Sham potential. $\Delta_{\mathbf{q}\nu}$ means the (first order) changes of the potential due to phonon $\mathbf{q}$ of branch $\nu$. 

In short, { a prerequisite for a large EPC  are large matrix elements $g_{mn\nu}$. This is particularly relevant in the case of LK-99, where only two electronic bands are relevant. They correspond to the $d_{xz}, d_{yz}$ orbitals \cite{griffin2023origin}. These orbitals are nearly degenerate (see Fig.~\ref{fig:fermisurface}), leading us to anticipate the most substantial effect when a phonon disrupts the degeneracy between these orbitals. Although the lifting of a degeneracy does not invariably result in a large EPC, it does occur in certain systems, such as MgB$_2$. In this system, lattice phonons that split the degeneracies along the $\Gamma-\mathrm{A}$ line are the most influential \cite{PhysRevLett.87.037001,C9TC02095K}. It is important to note, however, that MgB$_2$ and LK-99 are very different systems \cite{PICKETT2003117}, and this example is used solely to highlight the potential impact of lifting a degeneracy on $g_{nm\nu}$. This mechanism also has been used to explain the larger critical temperature of YB$_6$ with respect to nearly identical LaB$_6$ \cite{Yb6}.}

\begin{figure}[h]
    \centering
    \includegraphics[width=\columnwidth]{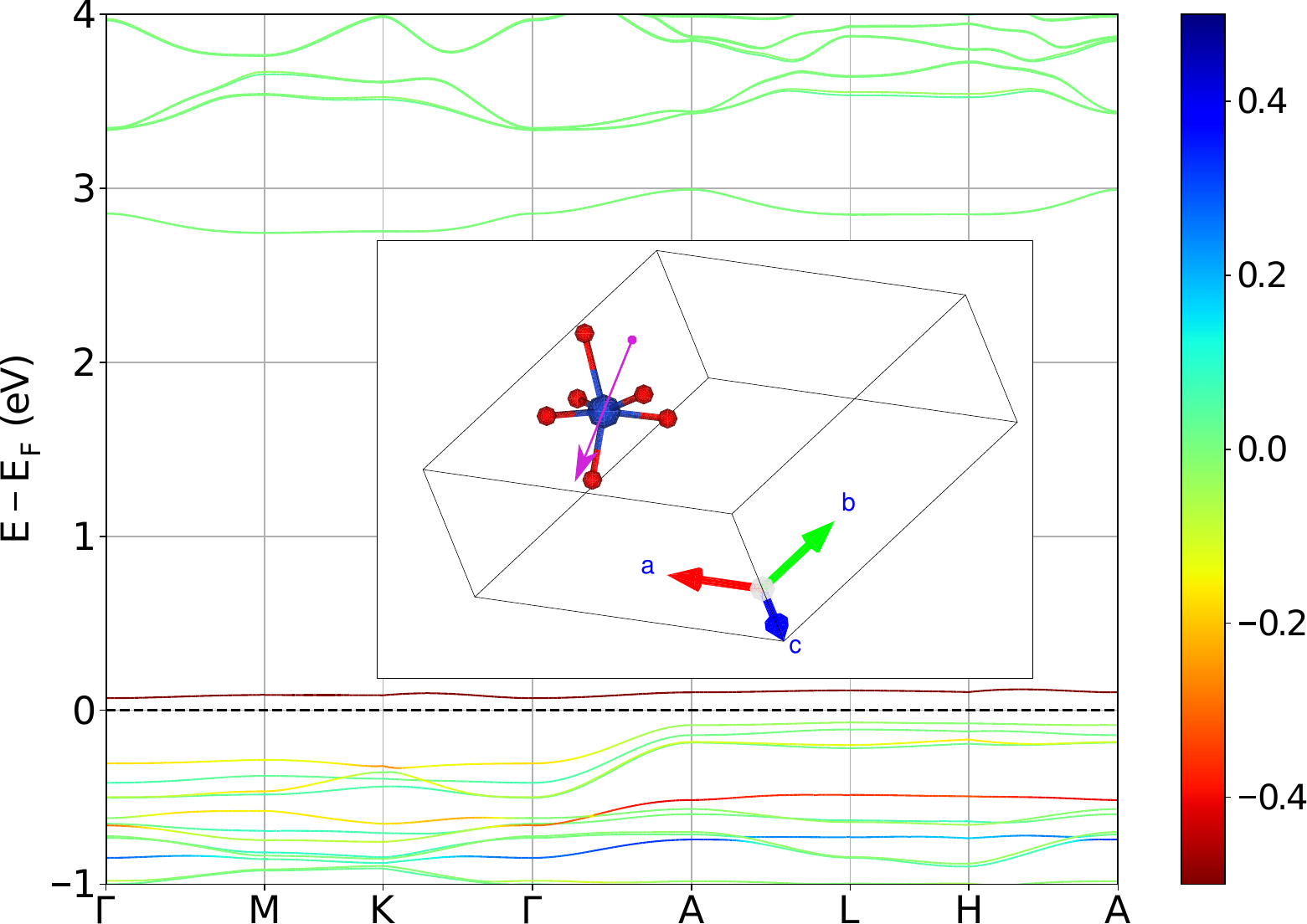}
    \caption{Band structure of LK-99 under the distortion from a frozen phonon. The color scale shows the projection of the wave function into the Cu atom, the positive (negative) values denotes the majority (minority) spin.  The inset shows the actual phonon at $\Gamma$ with a purple arrow. It is mostly localized in the Cu atom (blue). The d orbitals are no longer degenerate, with one of them buried $\sim 0.5$ eV below $E_F$ (red-orange band). The amplitude of the frozen phonon was 0.04\AA{}, mostly involving a shift of the Cu atom from its equilibrium position \cite{frozenphonon}.}
    \label{fig:epc}
\end{figure}

The effect of such a frozen phonon is dramatic, see Fig.~\ref{fig:epc}. Not only is the degeneracy lifted along $\Gamma-A$, but one of the d-bands is now buried $\sim 0.5$ eV below $E_F$ for every point of the reciprocal space. These changes notably turn LK-99 into an insulator. We are unaware of any similar effect in other materials. It is worth commenting on a similar approach involving frozen phonons, with a larger amplitude, which has been used in MgB$_2$ (see Fig. 3 of Ref.~\cite{PhysRevLett.87.037001}) to shed light on the EPC due to the $E_{2g}$ phonons, and similarly to predict a BCS  superconductor with large EPC \cite{singh2022high}. In both cases, a splitting of degenerate bands of about $\sim 2$ eV was observed. However, they continued being metals.

\section{\label{sec:triclinic}LK-99: Triclinic lattice}

\subsection{\label{sec:tric-geo}Crystal Structure and phonons}
\begin{figure}
    \centering
    \includegraphics[width=0.8\columnwidth]{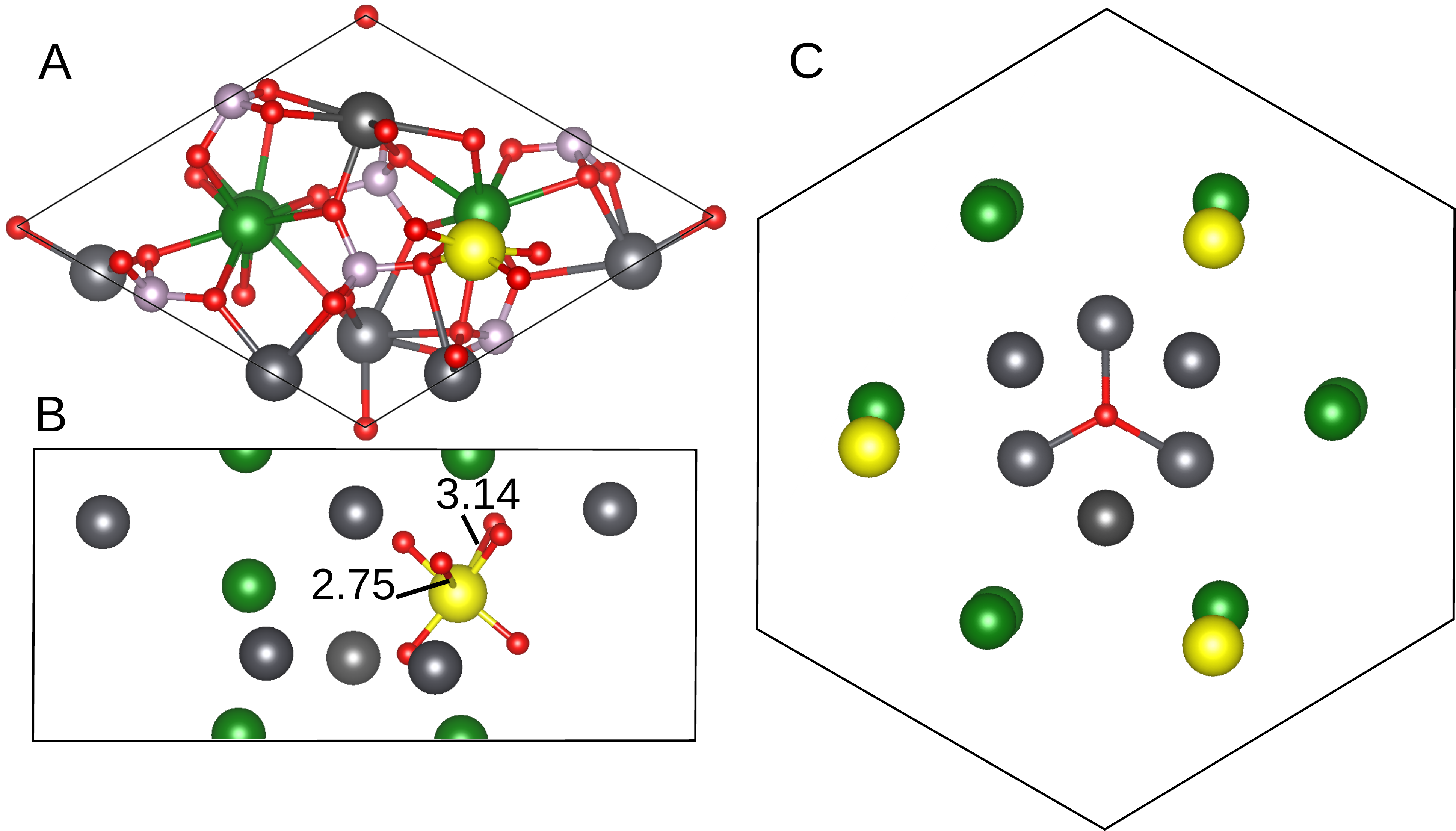}
    \caption{(a) Unit cell of the triclinic Pb$_{9}$Cu$_1$(PO$_4$)$_6$O obtained after relaxation. (b) Lateral view, with emphasis on the O atoms closer to Cu. The distances shown are in \AA{}, and the other distances are very close to 2.0\AA. (c) Top view removing atoms other than Cu, Pb, and the central O. The colors follow Fig.~\ref{fig:Pb-geo}, Cu is yellow.}
    \label{fig:triclinic}
\end{figure}

By following the distortion associated to one of the imaginary phonon branches (with zero momentum, $\Gamma$ point)  we obtained a crystal structure of Pb$_{9}$Cu(PO$_4$)$_6$O { stable at the harmonic level}, with triclinic lattice and space group \#1, see Fig.~\ref{fig:triclinic}. It is very close to a hexagonal lattice, see Fig.~\ref{fig:cu1-geo}(a) . Since the atomic rearrangements are centered in the Cu atom, we predict that \textit{(i)} following the same phonon with a different momentum, plus an adequate supercell, should result in different final geometries, \textit{(ii)} following the other imaginary phonon branch should yield a similar result to the one obtained, related by a symmetry operation. In a large system, both types of distortion could coexist, and studying them in a unit cell is unrealistic. The triclinic unit cell is 0.55 eV/f.u. lower in energy than the hexagonal. Even though this value may seem large, it is only 13 meV per atom. We do not study different magnetic arrangements in this section.

\begin{figure}
    \centering
    \includegraphics[width=\columnwidth]{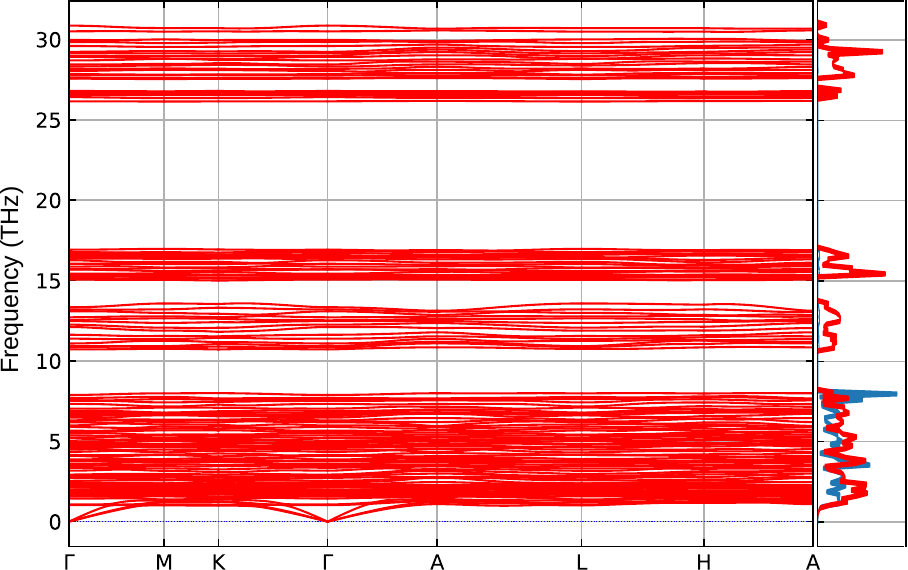}
    \caption{Phonon band structure for the triclinic LK-99-like crystal structure.{ The right panel is the total (red) and partial (Cu projection, gray) phonon density of states.}}
    \label{fig:phononstriclinic}
\end{figure}

The phonon band structure of the triclinic system has no imaginary branches, see Fig.~\ref{fig:phononstriclinic}. Otherwise, it shows the same features found in the hexagonal lattice.

\subsection{\label{tric-electron}Electronic Structure}

\begin{figure}
    \centering
    \includegraphics[width=\columnwidth]{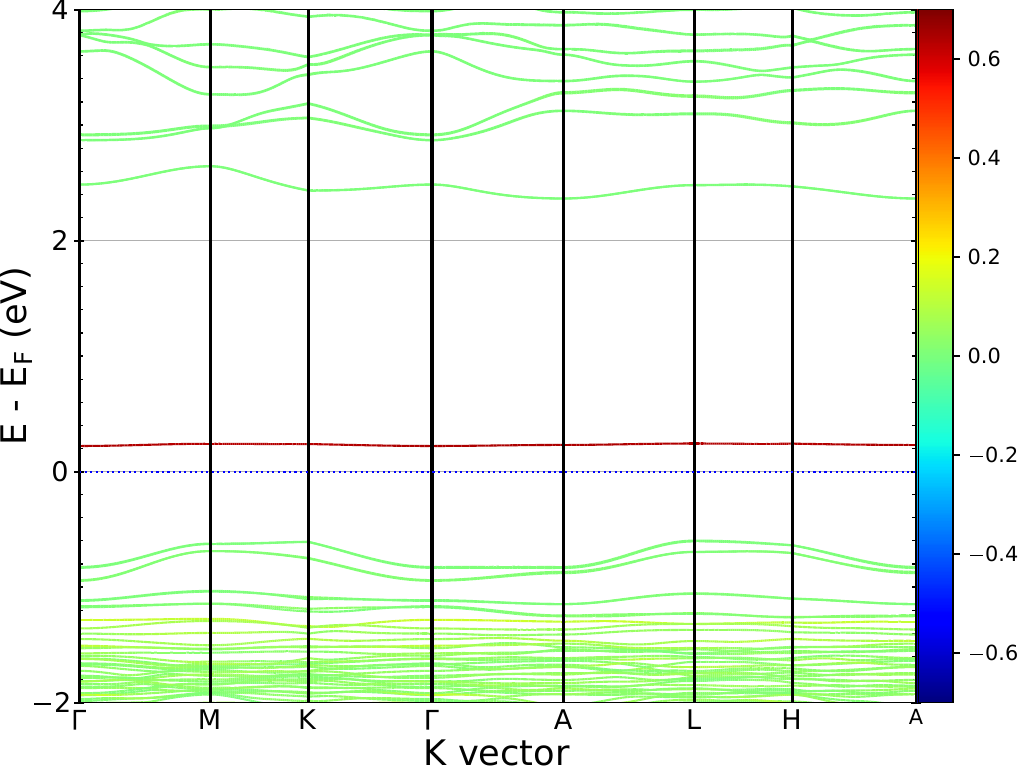}
    \caption{Band structure of the LK-99-like system with triclinic lattice. The color scale shows the projection of the wave function into the Cu atom, the positive (negative) values denotes the majority (minority) spin. There is a single band within the fundamental band gap; it is not spin-degenerate.}
    \label{fig:bands-triclinic}
\end{figure}

 The triclinic system's band structure (Fig.~\ref{fig:bands-triclinic}) resembles that of our frozen phonon calculation (Fig.~\ref{fig:epc}) and a recent study \cite{georgescu2023cudoped}. This is not surprising since in the hexagonal lattice, the degeneracy along $\Gamma$-A, plus the Fermi surface nesting suggest a Jahn-Teller-like distortion to lower the total energy \cite{hlinka2023possible}. 

Regarding superconductivity, the lowest energy structure of apatite-like Pb$_{9}$Cu(PO$_4$)$_6$O is an insulator. However, a small electron doping (\textit{e.g.} defects) should turn the system into a metal.



\section{\label{sec:discussion}Discussion}

We have studied two Cu-doped lead apatite systems, with hexagonal and triclinic lattice. The hexagonal lattice has promising electronic properties for superconductivity, but it is not stable against vibrations (at the harmonic level). The triclinic system has the lowest energy and is vibrationally stable; however, it is a band insulator. Now, we are in a position to comment on two relevant questions.

\subsection{What is the actual crystal structure of LK-99?}
\label{sec:what-crystal}

Does the LK-99 material assume the hexagonal or the triclinic lattice? We are convinced the picture is not so simple, even assuming the simplest stoichiometry, Pb$_{9}$Cu(PO$_4$)$_6$O. Even at low temperatures of $\sim 60$ K, the system should fluctuate between the different symmetry-related distortions, resulting in a larger effective symmetry than the triclinic system. { In our study we neglected anharmonic contributions, but they could be able to stabilize the hexagonal structure at high-enough temperature. Indeed, during the revision of this article, a preprint showed that anharmonic effects remove the imaginary phonon branches of the hexagonal structure at room temperature \cite{Kim_2024}.} { In the Appendix~\ref{sec:xdr}, we calculated the averaged XRD pattern along a molecular dynamics simulation, which agrees with a thermal stabilization of the hexagonal lattice.}

{On top of the previous discussion, we need to mention that up to now, we are unaware of any successful replication of the crystal, at least as proposed in the original preprints. Experimental samples suffer from Cu-clustering \cite{puphal2023single}, as evidenced by an inhomogeneous magnetism\cite{Liu2023phases}.}

\subsection{Is LK-99 a superconductor?}

{

From the previous discussion, a BCS superconducting state is unclear. Unless the crystal has a perfect hexagonal lattice, the system should be a band insulator. Let us assume the flat band of Fig.~\ref{fig:bands-triclinic} is half-occupied (which it is not); the EPC should not induce large changes in the electronic structure. The band is flat (atomic-like) and should remain so, unless another Cu d orbital gets close in energy, returning to something like the hexagonal crystal (\textit{i.e.} nearly degenerate half-filled bands), see Fig.~\ref{fig:bands-Cu-FM}. In BCS superconductors, usually, the EPC splits degenerate bands instead of inducing degeneracies. 

Even if the hexagonal lattice is stable at room temperature \cite{Kim_2024}, its electronic structure at room temperature could be different from the two-flat bands picture of Figs.~\ref{fig:bands-Cu-FM}, \ref{fig:fermisurface}. To get some insight into the electronic structure of this material at room temperature, we carried out a DFT-based molecular dynamics (MD) simulation of one unit cell. Fig.~\ref{fig:bandsMD} shows the energy levels as a function of the time. Since the band(s) of interest is (are) very flat, we only plotted the values at $\Gamma$. In the figure, the bands are colored according to their projection into Cu atoms. Along most of the dynamics, only one Cu-like level exists close to the Fermi level. The next Cu level lies about 1 eV below the Fermi level and it is hybridized with other atoms (pale purple stripe). Only at $t\approx 3500$ fs two Cu bands are at the Fermi level, making it an infrequent event. Despite the finite size errors from our MD simulations, we are confident that the ``two flat-bands picture'' does not hold at room temperature. Then we can discard the system as a high-temperature superconductor, even if doped to be metallic. Of course, if the experimental crystal structure differs from the one calculated here (\textit{e.g.} Cu clustering, Cu atoms in nonequivalent lattice positions) this analysis would need to be reassessed.

\begin{figure}
    \centering
    \includegraphics[width=\columnwidth]{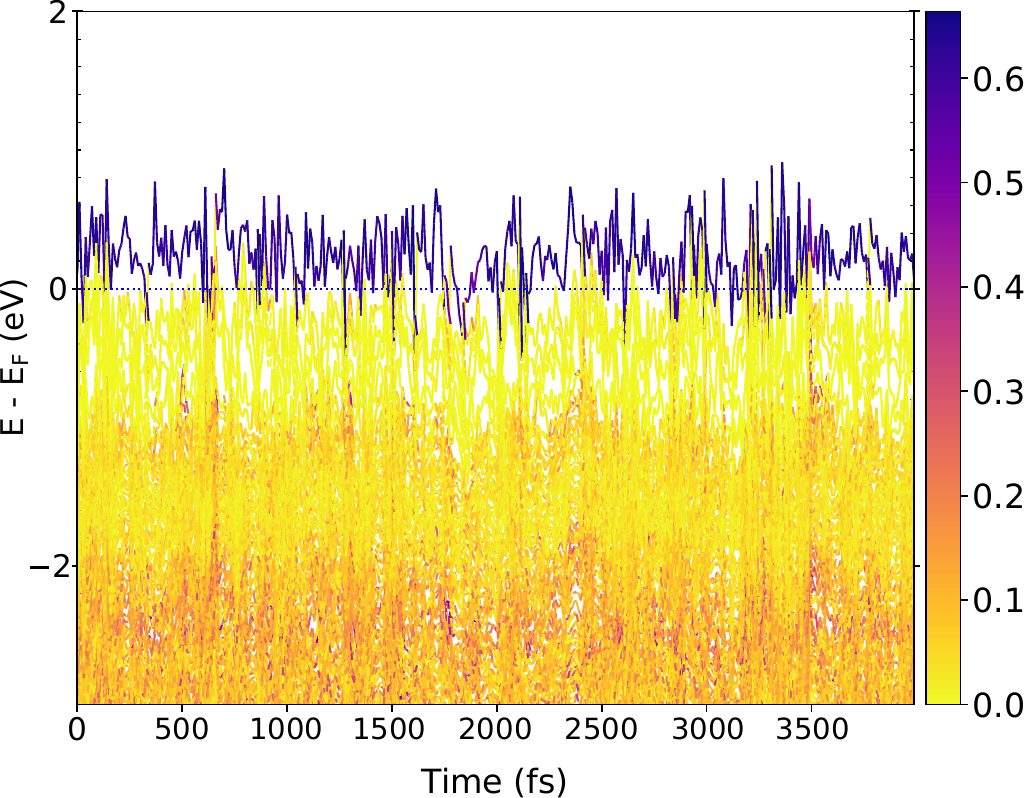}
    \caption{{Energy levels at $\Gamma$ of a unit cell of Pb$_{9}$Cu(PO$_4$)$_6$O at room temperature. The color intensity is the projection of each wave function on Cu atoms.}}
    \label{fig:bandsMD}
\end{figure}

}

\section{\label{sec:con}Conclusions}

The so-called LK-99 material is a Cu-doped lead oxyapatite system. We studied specifically the system Pb$_9$Cu(PO$_4$)$_6$O, with two possible crystal structures: a hexagonal and a triclinic lattice. 

The hexagonal lattice has attracted much attention, and it features practically flat, half-filled bands. They are spin-split. The Fermi surfaces show a nesting around $\Gamma$ and A. These bands are unstable against vibrations (\textit{i.e.} imaginary phonon frequencies) { at least at the harmonic level}.

By following one of the imaginary phonon branches, we found a very similar triclinic lattice. It is lower in energy and vibrationally stable. However, it is a band insulator, with a single spin-split unoccupied band within the fundamental band gap. 

Since the energy barrier between both lattices is much smaller than room temperature, in a crystal the Cu atom should oscillate between the different phases, giving an overall symmetry larger than triclinic. Along these oscillations insulating and metallic states could take place. { However, having half-filled almost flat bands is an infrequent event.} 

A superconductor is unlikely, at least from the BCS picture.  {If the material is a superconductor, its crystal structure probably differs from the one being reported.}

\section{\label{sec:methods}Computational Methods}
We employed DFT as implemented in the VASP
package \cite{kresse1993ab,kresse1996efficiency,kresse1996g,PhysRevB.49.14251} using the
projector augmented wave method \cite{blochl1994projector} and
Perdew-Burke-Ernzerhof (PBE) exchange-correlation (XC)
functional \cite{perdew1996generalized}. For the lead apatite we made some calculations with the HSE06 hybrid functional \cite{heyd2003hybrid,krukau2006influence,paier2006screened},  but since all relevant physics is captured by simpler methods we kept PBE across the manuscript. To ensure the completeness of the basis, we set the energy cutoff to 520~eV. In the structural optimization, the k-points grid was set to $3\times 3\times 4$. {The smearing used for k-points integration was 10 meV.}

For phonon calculations, we used a classical force-field created with machine learning of a DFT molecular dynamics simulation \cite{ml1,ml2}. We used a supercell of size $3 \times 3 \times 3$. The phonopy \cite{togo2015first,togo2023first,togo2023implementation} software was employed for the post-processing of phonons. { The mesh for the phonon density of states was set to $10\times 10 \times 10$. More details about the force field generation are in the Appendix~\ref{sec:mlff}.} 

PyProcar \cite{pyprocar,LANG2024109063} was employed for analyzing the electronic structure. The IFermi \cite{Ganose2021} software was used to plot Fermi surfaces. Unless otherwise stated in the text, the calculations used the GGA$+U$ approach in the rotationally invariant scheme of Dudarev et. al \cite{PhysRevB.57.1505}. We used a value of $U=2.0$ eV for the d-orbitals of Cu. Electronic band structures without this correction or with a different value are qualitatively similar{, see Appendix~\ref{sec:varying-u}}. 

  To build the crystal structure of LK-99, we used as a starting point the crystal structure of apatite available on the website of the Materials project.\cite{materialsproject} Then the composition was adjusted to the one of LK-99, and the cell was fully relaxed. Structures were built with VESTA \cite{momma2011vesta}.

\section*{ACKNOWLEDGEMENTS}
This research was funded by FONDECYT projects 1220366, 1231487, 1220715 and by the Center for the Development of Nanosciences and Nanotechnology, CEDENNA AFB 220001. JCE and NFB gratefully acknowledge ANID for their national doctorate's scholarship year 2023 number 21231429 and national master's scholarship year 2022 number 22220676, respectively. CC acknowledges ANID for the grant ECOS210019. FM is supported by  Conicyt PIA/Anillo ACT192023. ANA thanks the Brown Science Foundation Award. Powered@NLHPC: This research was partially supported by the supercomputing infrastructure of the NLHPC (ECM-02). RHL performed calculations on UCLA's IDRE Hoffman2 cluster and acknowledges support by the United States National Science Foundation Graduate Research Fellowship under Grant No. 2034835. The authors want to thank Pablo D\'iaz and Jirka Hlinka for the discussions about this subject.

\appendix
{
\section{Fermi Surface and Fermi Nesting }
\label{sec:nesting}
\begin{figure*}[htp]
    \centering
    \includegraphics[width=\textwidth]{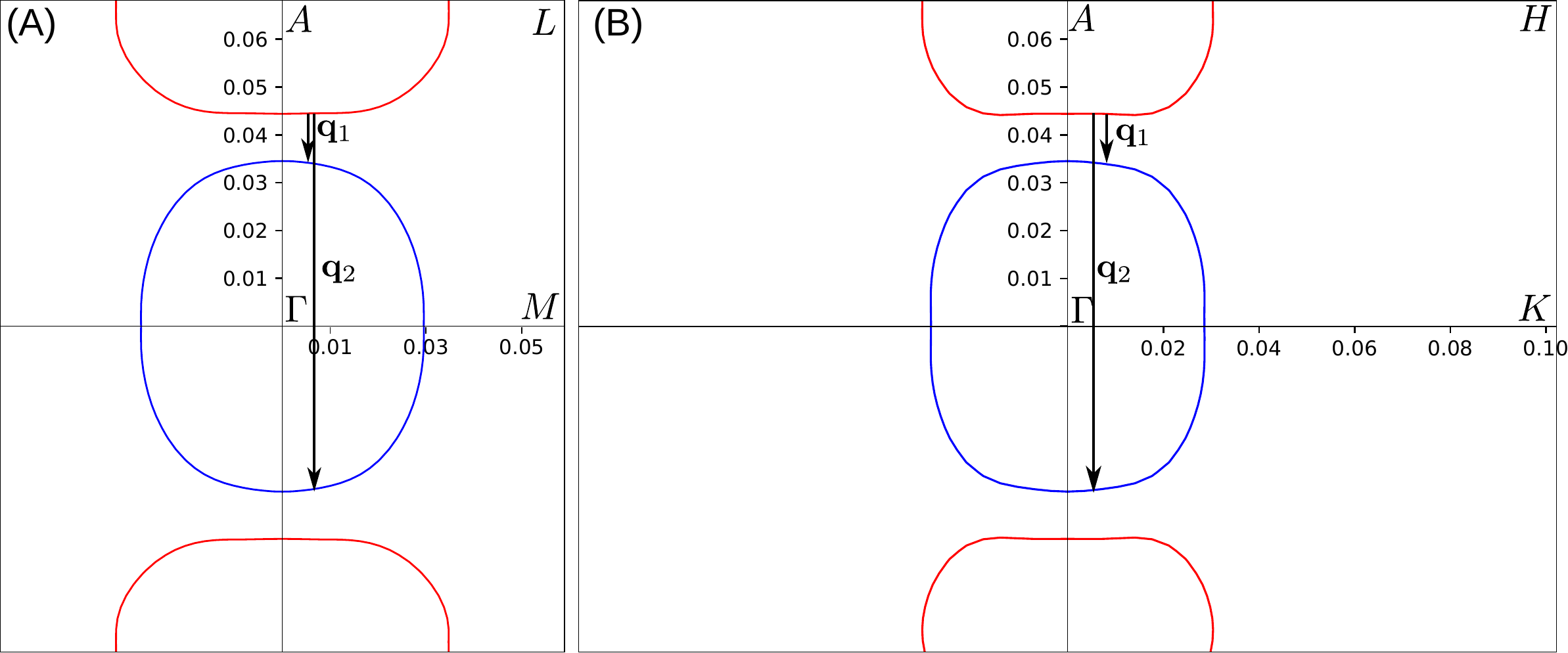}
    \caption{Two-dimensional contour plots of the Fermi surface of the hexagonal lattice of LK-99. The bands are those plotted in Fig.~\ref{fig:fermisurface}. Panels (A) and (B) show different planes. The arrows represent a pair of phonons with momentum $\mathbf{q}_1, \mathbf{q}_2$ connecting two Fermi surfaces. The length of the arrow is the same in both panels.}
    \label{fig:nesting}
\end{figure*}

The phonon self-energy can be written as:
\begin{align}
    \Pi(\mathbf{q},\omega) =\\-2 \sum_{\mathbf{k}, n, n'}\frac{f(\epsilon_{n,\mathbf{k}})-f(\epsilon_{n',\mathbf{k+q}})}{\epsilon_{n,\mathbf{k}}-\epsilon_{n',\mathbf{k+q}}-\hbar\omega -i\delta}M^{2}_{n\mathbf{k},n'\mathbf{k+q}},
\end{align}
where the band momentum and indexes are $\mathbf{k}, n, n'$,  respectively. The occupation of a given state with energy $\epsilon$ is given by $f(\epsilon)$. The phonon momentum and energy are $\omega$ and $\mathbf{q}$. $M_{n\mathbf{k},n'\mathbf{k'}}$ are the effective matrix elements of the deformation potential\cite{Pickett1977}. A divergence of the self-energy may arise at $\omega=0$, if for a given vector $\mathbf{q}$ the bands are degenerate in a large region of the reciprocal space; in other words, the Fermi surfaces are nested.

Due to the difficulty in calculating $M$, it is often approximated as a constant, and the imaginary part of the bare electronic susceptibility with $\omega\to0$, $\chi''_0(\mathbf{q})$,  is used to quantify the Fermi nesting\cite{nesting}:
\begin{equation}
    \lim_{\omega\to 0 }\frac{\chi''_0(\mathbf{q})}{\omega} = \sum_{n\mathbf{k}}\delta(\epsilon_{n,\mathbf{k}}-\epsilon_F)\delta(\epsilon_{n',\mathbf{k+q}}-\epsilon_F),
\end{equation}
with $\epsilon_F$ the Fermi energy. Although the calculation of the nesting can, in principle, be done with DFT \cite{epw,nesting}, our attempt was impractical for a system this large that needs a dense grid of k-points to describe its Fermi surface.

Instead of an actual calculation of $\chi''_0(\mathbf{q})$, in Fig.~\ref{fig:nesting}, we provide a section of the Fermi surface and the vectors $\mathbf{q}$ connecting two Fermi surfaces. Both Fermi surfaces have little dispersion near their crossing of the $\Gamma-A$ line, favoring an overlap of $\epsilon_{\mathbf{k}}$ and $\epsilon_{\mathbf{k+q}}$. 
}

{
\section{XRD}
\label{sec:xdr}

\begin{figure}
    \centering
    \includegraphics[width=\columnwidth]{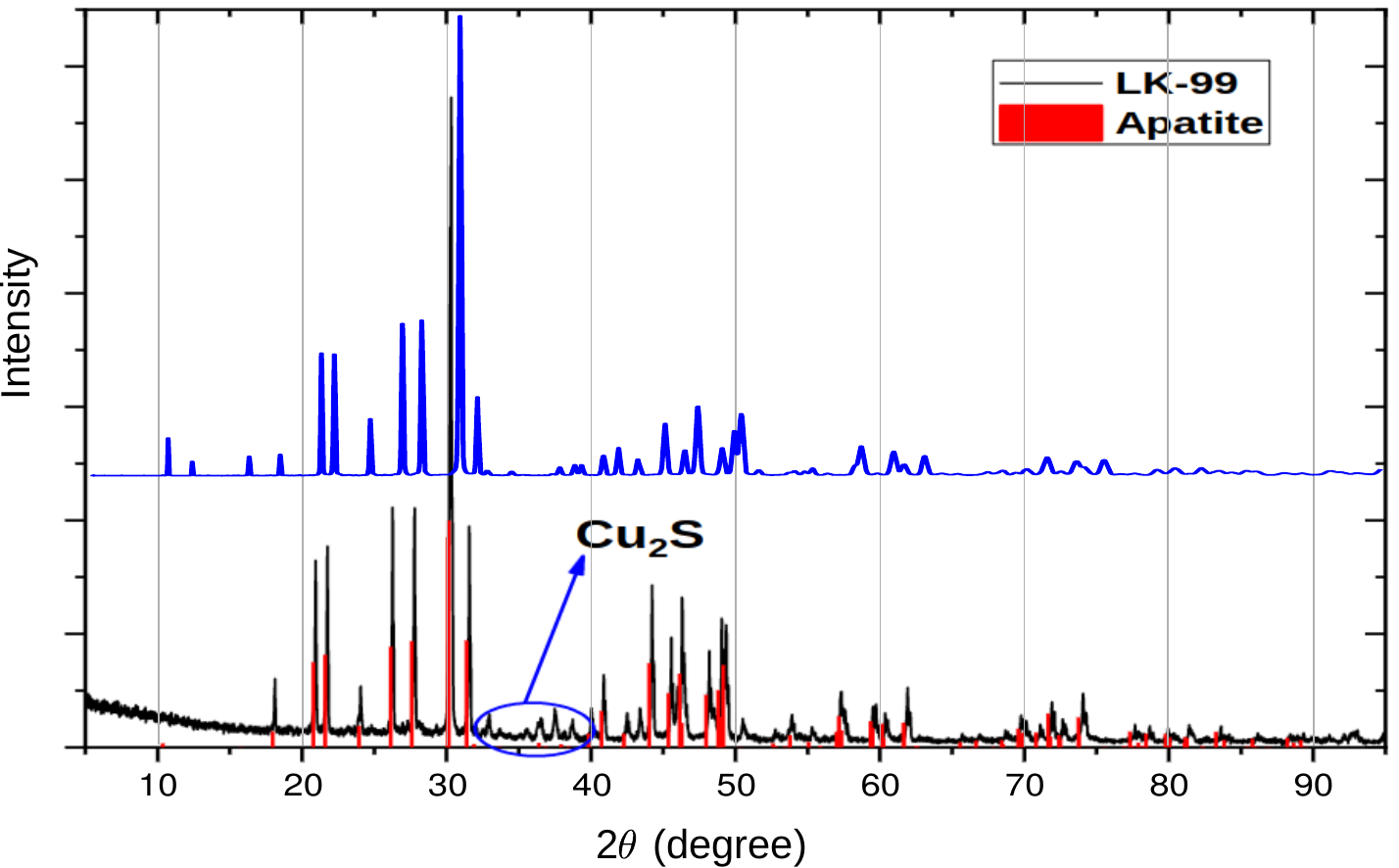}
    \caption{XRD of LK-99 obtained in this work (blue line). For comparison,  Figure 2 of \cite{lee2023} is included (black and red lines).}
    \label{fig:xrd}
\end{figure}

We ran a DFT-machine-learned classical molecular dynamics simulation of this system at room temperature, with a $3\times 3\times 3$ supercell ($\sim 1200$ atoms); see Fig.~\ref{fig:xrd}. The calculated XRD pattern corresponds to the average along the last 500 fs of the simulation. In general, this is in good agreement with the experimental results. However, our peaks are shifted to larger angles, which is explained by the thermal expansion of the lattice in our simulation. This picture agrees with a hexagonal lattice stabilized by temperature\cite{Kim_2024}.

}

\section{Machine Learning Force Fields}
\label{sec:mlff}
\begin{figure}
    \centering
    \includegraphics[width=\columnwidth]{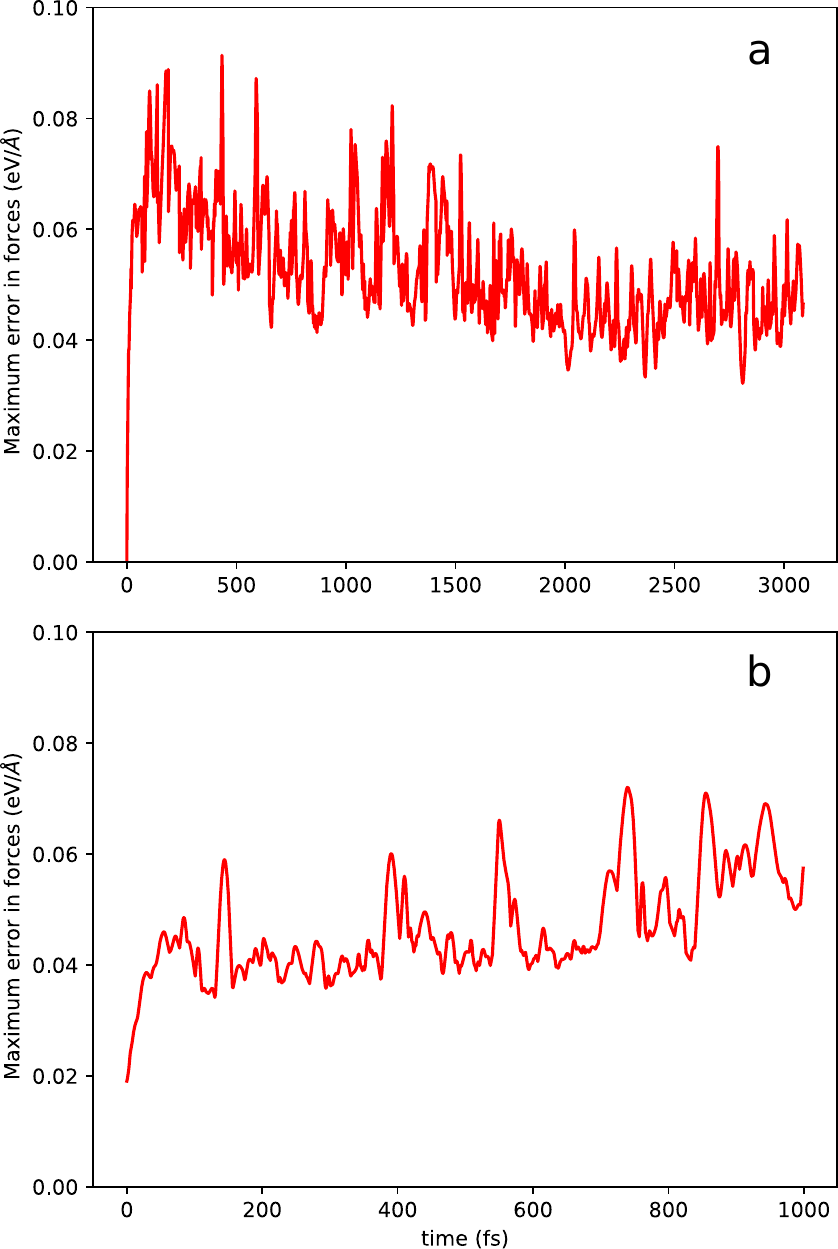}
    \caption{Bayesian estimation of the error on the forces, or statistical uncertainty. Only the largest error for each time step is plotted. (a) Fitting the force field by mixing DFT and the force field, the figure corresponds to the last refinement of the force field. (b) Molecular dynamics of a $3\times 3\times 3$ supercell at room temperature, only using the force field.}
    \label{fig:beef}
\end{figure}

{The training of a force field was performed on-the-fly: if the statistical uncertainty of the forces is above a certain threshold, the calculation is made with DFT, otherwise, the force field is used. This allows the use of DFT only as it is needed (\textit{i.e.} new configurations outside the validity domain of the force field). The threshold to switch between DFT or the force field is defined on the fly to refine the force field accuracy along the calculation. 

We started the training with a fixed lattice and low-temperature. Then we  allowed the lattice to relax to improve the force field. Finally, we increased the temperature to 400 K. Fig.~\ref{fig:beef}a shows the convergence of the Bayesian error in the forces for this last stage. For the simulation of a larger cell, we only used the force field; the statistical error in forces lies close to the training set, Fig.~\ref{fig:beef}b. It is worth noting that this level of error provides a qualitatively correct picture. For instance, the force field allowed us to identify the triclinic structure as the minimum energy. A posterior relaxation with DFT of the triclinic structure resulted in minimal changes. 
}

\section{Projected Density of States}
\label{sec:PDOS}
\begin{figure}[h]
    \centering
    \includegraphics[width=\columnwidth]{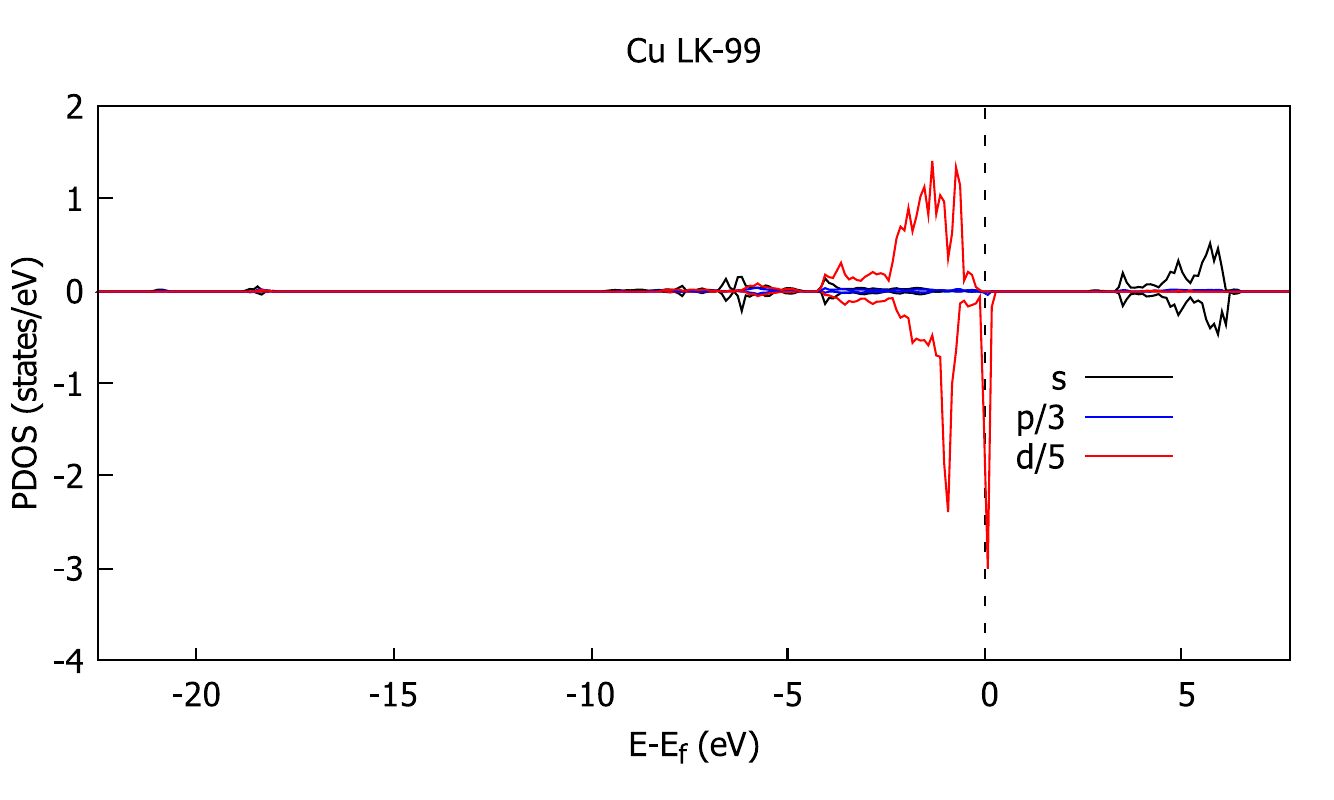}
    \caption{Projected density of states on Cu atom in LK-99 with stacking A and an FM ground state. Labels $p/3$ and $d/5$ indicate that the sum of PDOS is divided by 3 and 5, respectively.}
    \label{fig:DPOS_Cu_LK99_sum}
\end{figure}

{We used the VASPKIT code \cite{WANG2021108033} to obtain the projected density of states (PDOS) on Cu atom in LK-99 with stacking A and an FM ground state from VASP calculated data.  Fig. \ref{fig:DPOS_Cu_LK99_sum} shows the PDOS of the s, p, and d orbitals of Cu atom.}

\section{Dependency of the value of $U$}
\label{sec:varying-u}
\begin{figure*}
    \centering
    \includegraphics[width=0.9\textwidth]{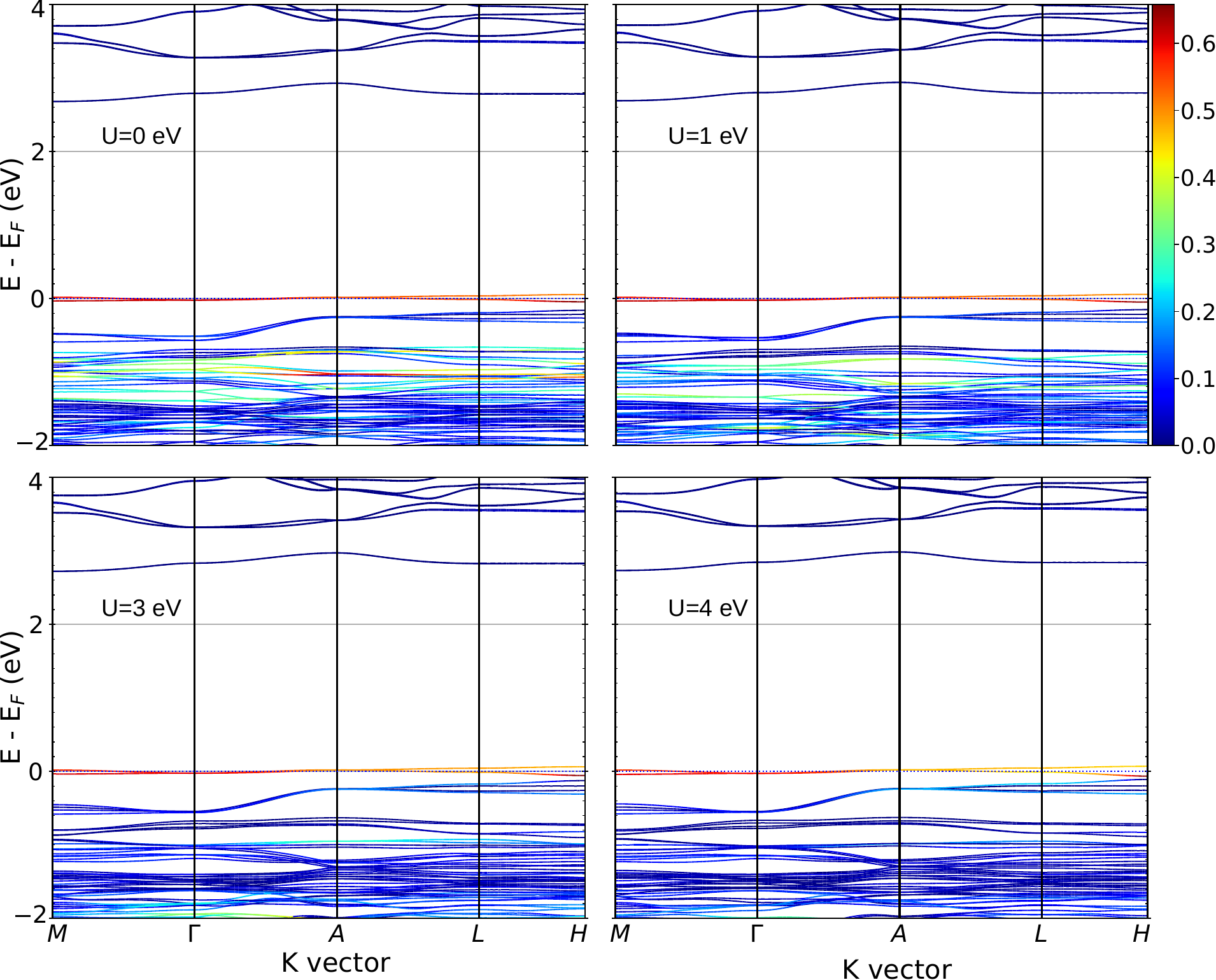}
    \caption{Band structure of the hexagonal lattice phase of LK-99 of some values of the Hubbard-like term $U$. The color bar denotes the projection of Cu atoms to the state.}
    \label{fig:u-bands}
\end{figure*}

\begin{figure*}
    \centering
    \includegraphics[width=\textwidth]{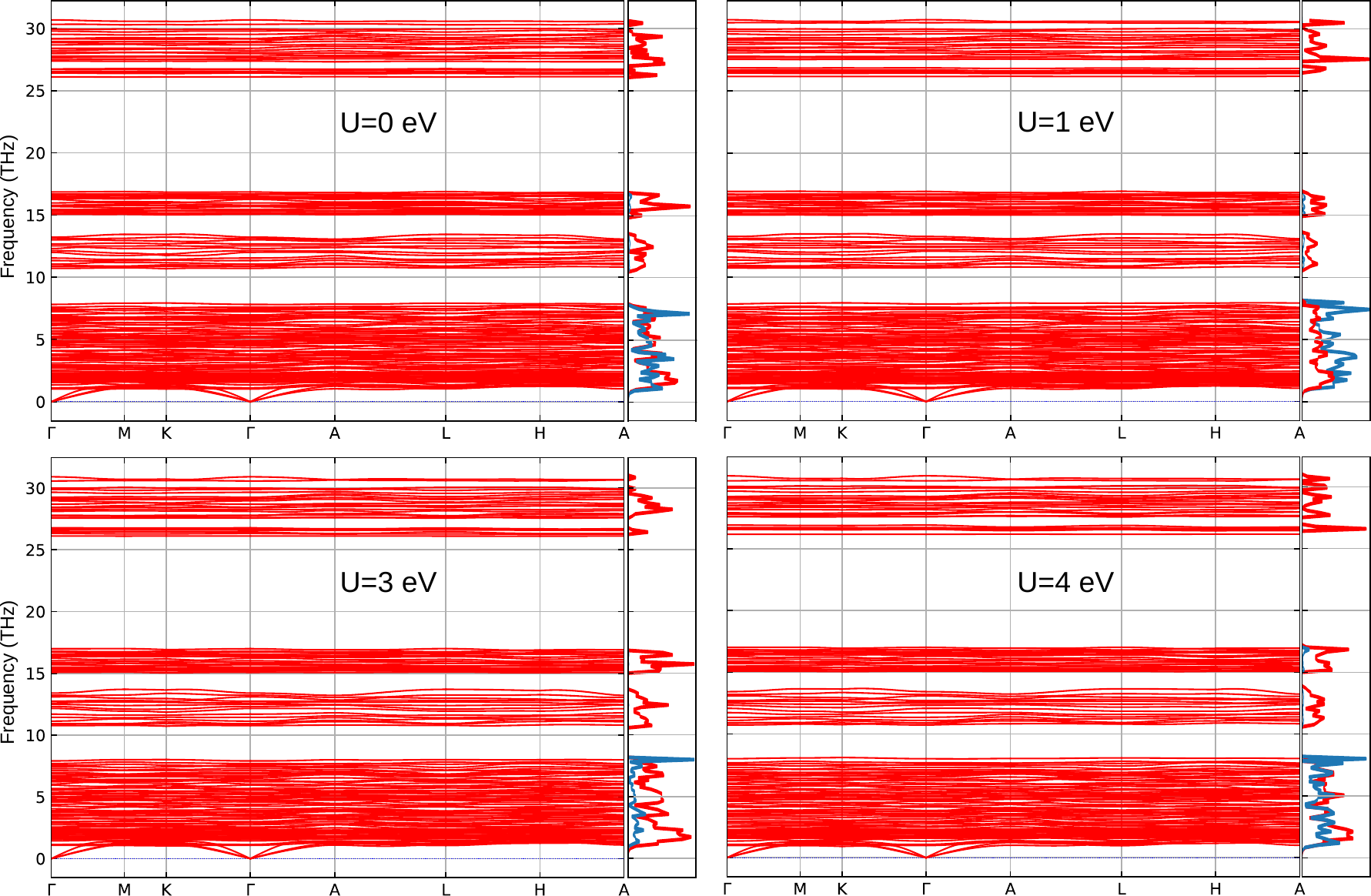}
    \caption{Phonon band structure of the optimized geometry for different values of the Hubbard-like term $U$. The Total (red) and partial (projected on Cu, grey) phonon density of states is at the left of each panel, both are in different scales. }
    \label{fig:u-phonons-triclinic}
\end{figure*}

{To explore the dependency of the calculations with respect to the Hubbard-like term $U$ we calculated the band structure of the system in the hexagonal lattice with $U=0-4$ eV, see Fig.~\ref{fig:u-bands}. The changes close to the Fermi energy are minimal. However, the participation of Cu in the valence band ($\sim -1$ eV) decreases when increasing the value of $U$. 

The phonons of the triclinic lattice, see Fig.~\ref{fig:u-phonons-triclinic}, are practically identical regardless of the value of $U$. The geometry was fully optimized for each $U$. The changes in the phonon density of states are likely due to the mesh size. 

}

\bibliography{apssamp}

\begin{thebibliography}{85}
\bibitem [{fro()}]{frozenphonon}%
  \BibitemOpen
  \href@noop {} {}\bibinfo {note} {To get a phonon involving the Cu
  displacement we calculate the vibrational modes of the Cu and its neighbor O
  atoms, while keeping all other atoms fixed. We chose the mode with the
  largest Cu displacement to apply a frozen phonon. While the vibrational mode
  obtained this way is not any actual phonon, it can be expanded into the basis
  of the actual phonons of the system.}\BibitemShut {Stop}%
\end{thebibliography}
\end{document}